\documentclass[useAMS,usenatbib,usegraphicx]{mn2e}

\usepackage{rotating,amssymb}

\def\hd{HD\,191612}
\def\l{$\lambda$}
\def\ha{H$\alpha$}
\def\hb{H$\beta$}
\def\he{He\,{\sc i}}
\def\hee{He\,{\sc ii}}
\def\niii{N\,{\sc{iii}}}

\def\ciii{C\,{\sc{iii}}}
\def\civ{C\,{\sc{iv}}}

\def\kms{km\,s$^{-1}$}

\def\xmm{{\sc XMM}\emph{-Newton}}
\def\ros{\emph{{\sc ROSAT}}}
\def\th{$\theta^1$~Ori~C}

\title[Towards an understanding of the Of?p star \hd ]{Towards an understanding of the Of?p star \hd: \\ Phase-resolved multiwavelength observations\thanks{Based on observations collected at the Haute-Provence Observatory (France) and with \xmm, an ESA Science Mission with instruments and contributions directly funded by ESA Member States and the USA (NASA).}}

\author[Y. Naz\'e et al.]{
Ya\"el Naz\'e$^{1}$\thanks{E-mail: naze@astro.ulg.ac.be}\thanks{Post-doctoral Researcher FNRS (Belgium)}, Gregor Rauw$^1$\thanks{Research Associate FNRS (Belgium)}, A.M.T. Pollock$^2$, Nolan R. Walborn$^3$, \and Ian D. Howarth$^4$\\
$^{1}$ Institut d'Astrophysique et de G\'eophysique, Universit\'e de Li\`ege, B\^at. B5c,  All\'ee du VI Ao\^ut 17, B-4000 Li\`ege, Belgium.\\
$^2$ European Space Agency, \xmm\ Science Operations Centre, European Space Astronomy Centre, \\Apartado 50727, Villafranca del Castillo, 28080 Madrid, Spain\\
$^3$ Space Telescope Science Institute\thanks{Operated by the Association of Universities for Research in Astronomy, Inc., under NASA contract NAS5-26555.}, 3700 San Martin Drive, Baltimore, MD 21218, USA\\
$^4$ Department of Physics and Astronomy, University College London, Gower Street, London WC1E6BT, UK\\
}

\begin{document}

%\date{Accepted ??. Received ??; in original form ??}

\pagerange{\pageref{firstpage}--\pageref{lastpage}} \pubyear{2006}

\maketitle

\label{firstpage}

\begin{abstract}
We present the analysis of phase-resolved X-ray and optical observations of the peculiar hot star \hd\ (Of?p). This star is known to display line-profile variations that are recurrent with a period of 538\,d and its spectrum was found to present the signature of a magnetic field. In the X-rays, it is slightly overluminous compared to the canonical $L_{\rm X}/L_{\rm BOL})$ relation and appears brighter when the optical lines are strongest. Our \xmm\ observations further reveal that the X-ray spectrum of \hd\ exhibits rather broad lines and is dominated by a `cool' (0.2-0.6\,keV) thermal component, two characteristics at odds with the proposed magnetic rotator model.  We also report for the first time the low-level variability of the  metallic (abs/em) lines and \hee\ absorptions that appear to be associated with radial-velocity shifts. Finally, we compare our results with observations of the early-type stars and discuss several possible scenarios.

\end{abstract}


\section{Introduction}

In 1972, a new subtype of massive stars, the Of?p category, was defined \citep{wal72}. It gathers stars that present \ciii\ emission lines around 4650\,\AA\ of an intensity comparable to that of the neighbouring \niii\ lines. At that time, only two stars belonged to this category, HD\,108 and HD\,148937, but a third one was soon added, \hd\ \citep{wal73}.

Of the Of?p stars, HD\,108 has been the most studied in the past, but conflicting results were reported by different authors. To attempt to resolve the confusing situation, a long-term observing campaign was undertaken at the Haute-Provence Observatory \citep{naz01}. A detailed analysis of the data allowed most of the older models to be discarded and showed that the star displays line-profile variations in the Balmer and the \he\ lines. \citet{naz01} also discovered that these line-profile variations seem to be recurrent with a timescale of approximately 50--60 years (see also \citealt{naz06}). 

Soon after, \citet{wal03} reinvestigated another Of?p star, \hd, and discovered a similar phenomenon. Broad-band photometry of \hd\ had been shown to exhibit a clear modulation with a period of 538\,d \citep{koe02,naz04} and comparison with the spectroscopic results indicated that the spectral variations and photometric changes are correlated: when the star is faint, its emission lines are weakest and the star presents an apparent O8 spectral type, whereas when it is brighter, the emission lines are at maximum and the spectral type is O6.5 \citep{wal04}. 

Recently, \citet{donati} reported the detection of a magnetic field through the analysis of the Zeeman signature in several lines of \hd, notably the variable \he\ lines. They proposed the star to be an evolved version of \th, the only O-type star for which a magnetic field had been convincingly measured until then. \citet{donati} suggested that the 538\,d period of \hd\ was related to the rotation period of the star, which would have been braked as a result of the intense magnetic field. Unfortunately, neither \citet{donati} nor other authors have yet observed a magnetic signature in HD\,108 or modelled such a braking for this potential ``twin'' of \hd.

To understand better the nature of Of?p stars, a detailed study of their high-energy properties is needed. For HD\,108, observations were taken with \xmm\ in August 2002, revealing a thermal X-ray spectrum \citep{hd108xmm}. The star also appears slightly overluminous in X-rays. Moreover, its X-ray luminosity was not significantly different from the previous detections by $Einstein$ and $ROSAT$ (noting, however, that the old data have very large error bars and that a decrease in flux by less than a factor of 2 could not be excluded). These X-ray observations ruled out a compact-companion scenario where the wind ionization structure would be affected by the varying X-ray emission from an accretion disk \citep{hd108xmm}.

The shorter and better-known period of \hd\ allows us to study its X-ray properties at crucial phases of the 538\,d cycle. Two years ago, a large observing campaign was therefore initiated to get multiwavelength, phase-resolved observations of \hd. We obtained four \xmm\ pointings (PI: Naz\'e) and many visible echelle spectra with the 1.93\,m telescope of the Haute-Provence Observatory (PI: Rauw). We present in this paper an analysis of these two datasets. Additional optical data will be reported by Howarth et al. (in prep.), and throughout this paper, we will use their ephemeris: $T_0$(HJD)=2\,453\,415.2$\pm$0.6, $P$=537.7$\pm$0.5d, where $\phi$=0 corresponds to the maximum of the \ha\ emission (i.e. a phase shift of 0.5 compared to the ephemeris defined in \citealt{wal04}).

This paper is organized as follows. Section 2 describes the observations while Sections 3 and 4 present the results of the X-ray and optical OHP campaign, respectively. These observations are further discussed in Sect. 5 and we finally conclude in Sect. 6.

\section{Observations}
\subsection{Haute-Provence Observatory}

Spectroscopic observations of \hd\ in the visible domain were gathered during two years at the Haute-Provence Observatory (OHP). Most of the data  were obtained with the 1.93\,m telescope equipped with the Elodie echelle spectrograph, which covers the  3850--6800\,\AA\ wavelength range at a resolving power of 42\,000. Exposure times between 30 and 60 minutes were used, and the signal-to-noise ratio was generally about 100 or more. The Aur\'elie spectrograph at the 1.52\,m telescope enabled us to get a few additional spectra. Three different settings were used: one covering the range 6360--6750\,\AA\ at a resolving power $R$ of 11000, the second 5480--5910\,\AA\ at $R$=9000, and the last 4450--4900\,\AA\ at $R$=7000. Typical exposure times were 15 minutes. All the reductions were performed using the {\sc midas} software developed at ESO. \\

\subsection{\xmm}

\hd\ was observed four times with \xmm\ for approximately 10--20 ks in 2005: Revolution (Rev.) 975 on Apr. 5, Rev. 981 on Apr. 17, Rev. 1004 on June 2 and Rev. 1068 on Oct. 8. For these observations, the three European Photon Imaging Cameras (EPICs) were operated in the standard, full-frame mode and a medium filter was used to reject optical light. The second observation (Rev. 981) was granted because the first data were affected by a rather long background flare; unfortunately, a long episode of high radiation took place at the time of this replacement observation, rendering the data very noisy. 

\begin{figure}
\centering
\includegraphics[width=7cm]{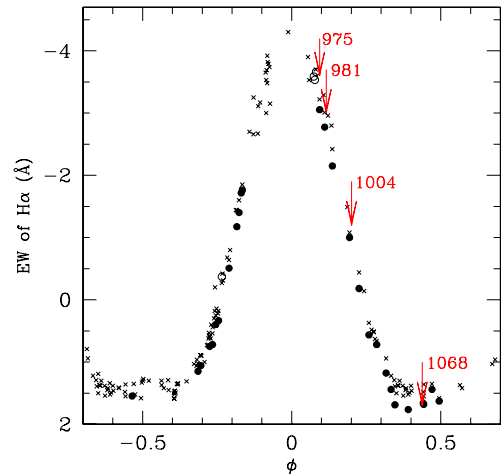}
\caption{The phases of the \xmm\ observations (between 2005 April and October) compared to the EW curve of the \ha\ line. Negative and positive EWs correspond to emission and absorption lines, respectively. The open circles indicate Aur\'elie data, filled symbols refer to Elodie observations, and crosses indicate other optical data (see Howarth et al., in preparation). \label{fig: haxmm}}
\end{figure}

We used the Science Analysis System (SAS) software, version~6.5, to reduce the X-ray data. They were first processed through the pipeline chains (tasks {\sc emproc, epproc} and {\sc rgsproc}), and then filtered in order to keep only the best data: as recommended by the SAS team, we kept for the EPIC MOS (Metal Oxide Semi-conductor) detectors single, double, triple and quadruple events (i.e. pattern between 0 and 12) that pass through the \#XMMEA\_EM filter; for the EPIC pn detector, only single and double events (i.e. pattern between 0 and 4) with flag$=$0 were considered. 

To check for contamination by low-energy protons, we have further examined the light curve at high energies (Pulse Invariant channel number$>$10000, E$\gtrsim$10\,keV, and with pattern$=$0). Where background flares were detected, we discarded the time intervals presenting high count rates (larger than 0.6~cts~s$^{-1}$ for EPIC MOS, 4.0~cts~s$^{-1}$ for EPIC pn, and 4--7.5~cts~s$^{-1}$ over the full energy range of the Reflection Grating Spectrometer, RGS). The best observations (i.e. Revs. 975, 1004, and 1068) were also combined with the tasks {\sc merge} for EPIC or {\sc rgscombine} for RGS. Further analysis was performed using the SAS, the FTOOLS tasks and the XSPEC software v 11.2.0. A table, available electronically at the CDS \footnote{http://cdsweb.u-strasbg.fr/}, provides the X-ray properties of all the sources detected in the field.

\section{X-ray Results}

\subsection{Light-curve}

The X-ray observations were planned to sample crucial phases in the 538\,d~period (Fig.~\ref{fig: haxmm}): the first observation was taken near the time of maximum emission in the visible domain while the last one was made during the `quiescent', low emission phase. All the X-ray data were taken during the same 538\,d cycle.

\begin{figure}
\centering
\includegraphics[width=8cm]{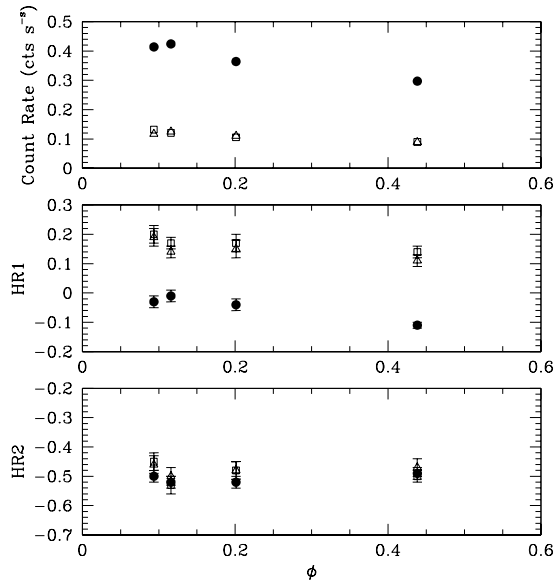}
\caption{\label{fig: cnt} Variations with phase of the count rate in the 0.4--10.0\,keV band (top) and of two hardness ratios (middle, bottom). Open symbols (triangles and squares) refer to EPIC MOS data, whereas filled circles represent EPIC pn observations. The hardness ratio $HR_1$ (resp. $HR_2$) is defined as $(M-S)/(M+S)$ (resp. $(H-M)/(H+M)$), where $S$ is the count rate in the 0.4--1.0\,keV band, $M$ in 1--2\,keV, and $H$ in 2--10\,keV.}
\end{figure}

The count rate of \hd, evaluated using the task {\sc edetect\_chain}, decreased by 40\% between the first (Apr. 5, Rev. 975, $\phi$=0.1) and the last (Oct. 8, Rev. 1068, $\phi$=0.4) observations, as shown In Fig.~\ref{fig: cnt}. At the same time, simultaneous visible observations taken at the OHP display a weakening of the emissions in the H and \he\ lines, as expected. From a detailed study of the behaviour in the visible of \hd\ \citep{wal04}, we know that these spectroscopic changes correspond to a decrease of the broad-band luminosity. Within the limitations of our dataset, our observations therefore suggest that the X-rays follow the behaviour observed in the visible range. Further data are however needed to fully assert this conclusion. 

Note that such a modest decrease in the X-rays could not be detected in the low signal-to-noise \ros\ data of Of?p stars; the result is thus consistent with the non-detection of such variations in HD\,108 \citep{hd108xmm}. Finally, on short timescales (i.e. during one pointing), we detected no significant variations of the X-ray emission. In fact, O stars generally lack variations on timescales this short (see, e.g., \citealt{san06}).

Looking at the hardness ratios of \hd\ (Fig.~\ref{fig: cnt}), two different behaviours appear, depending on the energy range considered: at lower energies (0.4--2.0\,keV), the decreasing X-ray emission clearly softens with time whereas the proportion of the harder emissions in the 1--2\,keV and 2--10\,keV ranges does not change significantly with phase. 

\subsection{Low-Dispersion Spectroscopy}

EPIC spectra of \hd\ were extracted over a circular region with radius 30\arcsec centred on the star. A faint source is present near \hd, so we did not choose an annular background region, but rather a nearby circle devoid of sources. For a colour excess $E(B-V)$ of 0.59 \citep{savage}, the interstellar absorbing column can be estimated as $N_{\rm int}^{\rm H}=3.4\times10^{21}$ cm$^{-2}$ \citep[using the relations of][]{boh78}. In the spectral modelling, we do not allow the absorption to go below this threshold.

\begin{figure*}
\centering
\includegraphics[width=8cm]{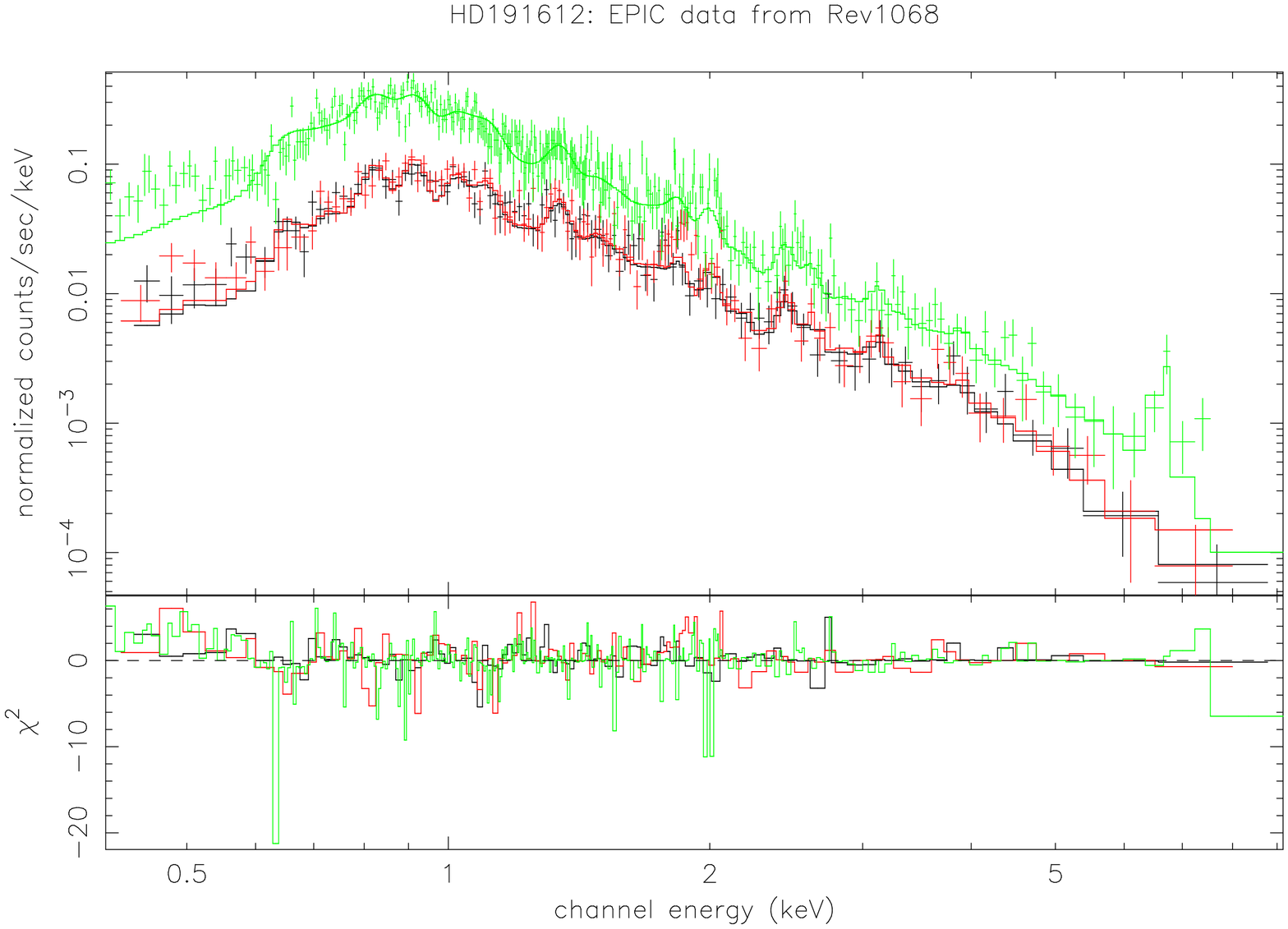}
\includegraphics[width=6cm]{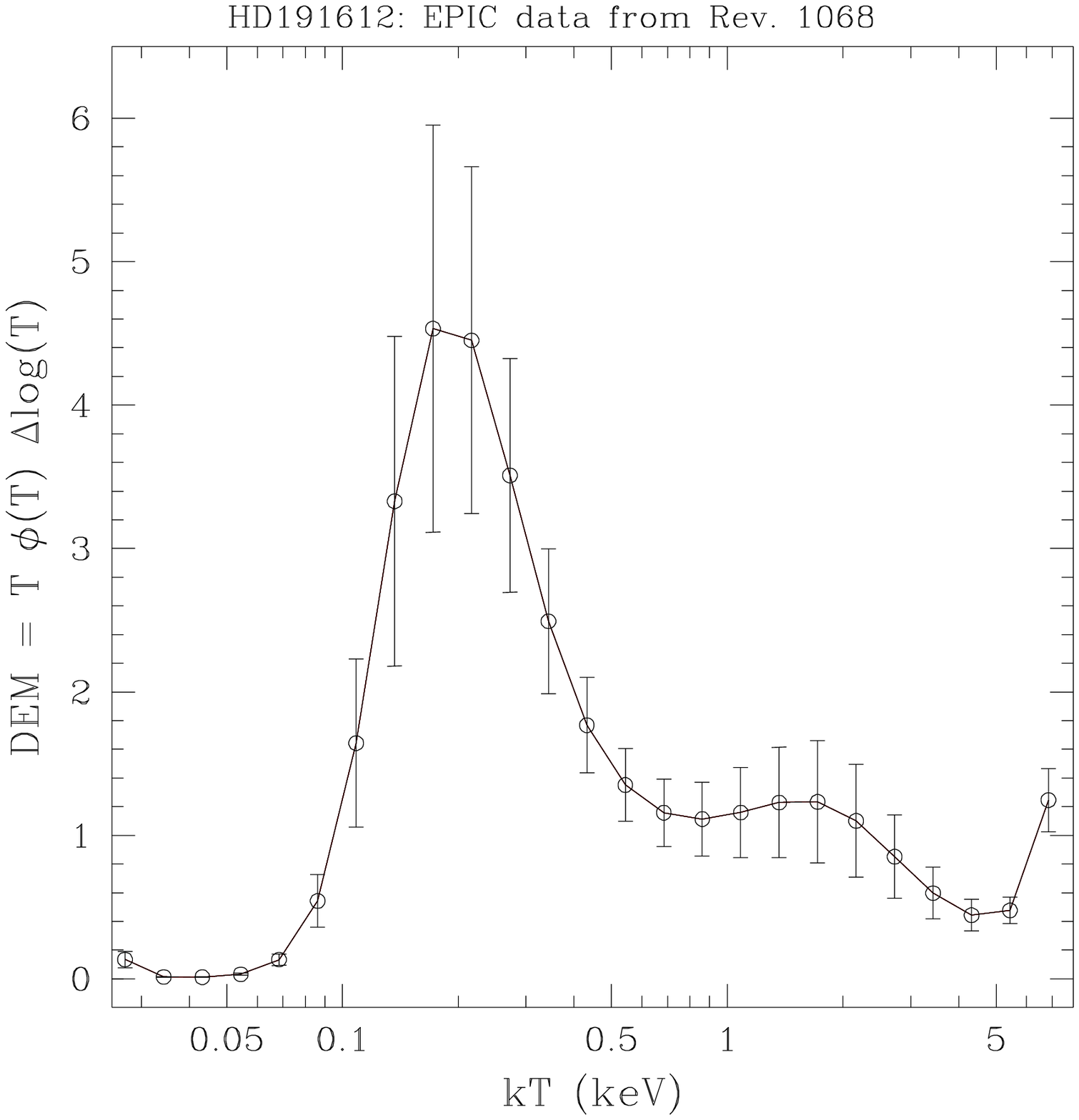}
\caption{{\it Left:}X-ray spectra of \hd\ recorded by EPIC-pn (upper data) and MOS (lower data) during Rev. 1068 ($\phi$=0.4, Oct. 2005), with the best-fit model superimposed (see Table~\ref{tab: table_fit} for values of the parameters). {\it Right:} Differential emission measure distribution from a fit of the same EPIC data by a DEM model where $\phi(T)$ corresponds to the exponential of a sum of Chebyshev polynomials between the 1st and 6th orders. The error bars were derived from Monte-Carlo simulations. \label{epicspec}}
\end{figure*}

The spectra from each instrument were fitted separately, but as they gave consistent results, within the errors, we finally fit all EPIC data simultaneously (Fig.~\ref{epicspec}). Fits using only one thermal component were unacceptable ($\chi^2>2$). We rather used the sum of two absorbed, optically thin equilibrium plasma models ($mekal$, \citealt{kaa92}). Since these two thermal components could arise in different regions of the wind, we allow them to be absorbed by independent column densities ($N^{\rm H}$). The best fit was obtained when the cooler component dominates, as is typically observed for `normal' O-type stars (see, e.g., \citealt{san06}). In fact, two sets of parameters gave similar values of the chi-squared (see Table \ref{tab: table_fit}), one with temperatures k$T$ around 0.6 and 1.0--2.5\,keV and negligible absorbing columns $N^{\rm H}$; the other with lower temperatures (0.24 and 1.0--1.6\,keV) and larger columns ($N^{\rm H}\sim 0.5$ and $1\times10^{22}$~cm$^{-2}$). The latter solution presents slightly lower $\chi^2$, even when including RGS data. Moreover, a fit of the EPIC data with a differential emission-measure model (DEM, $c6pmekl$, \citealt{lem,sin}) confirms the existence of a dominant temperature near 0.2\,keV whilst a secondary hump in the DEM distribution appears at 1--2\,keV (Fig.~\ref{epicspec}). 

\begin{figure}
\centering
\includegraphics[width=8cm]{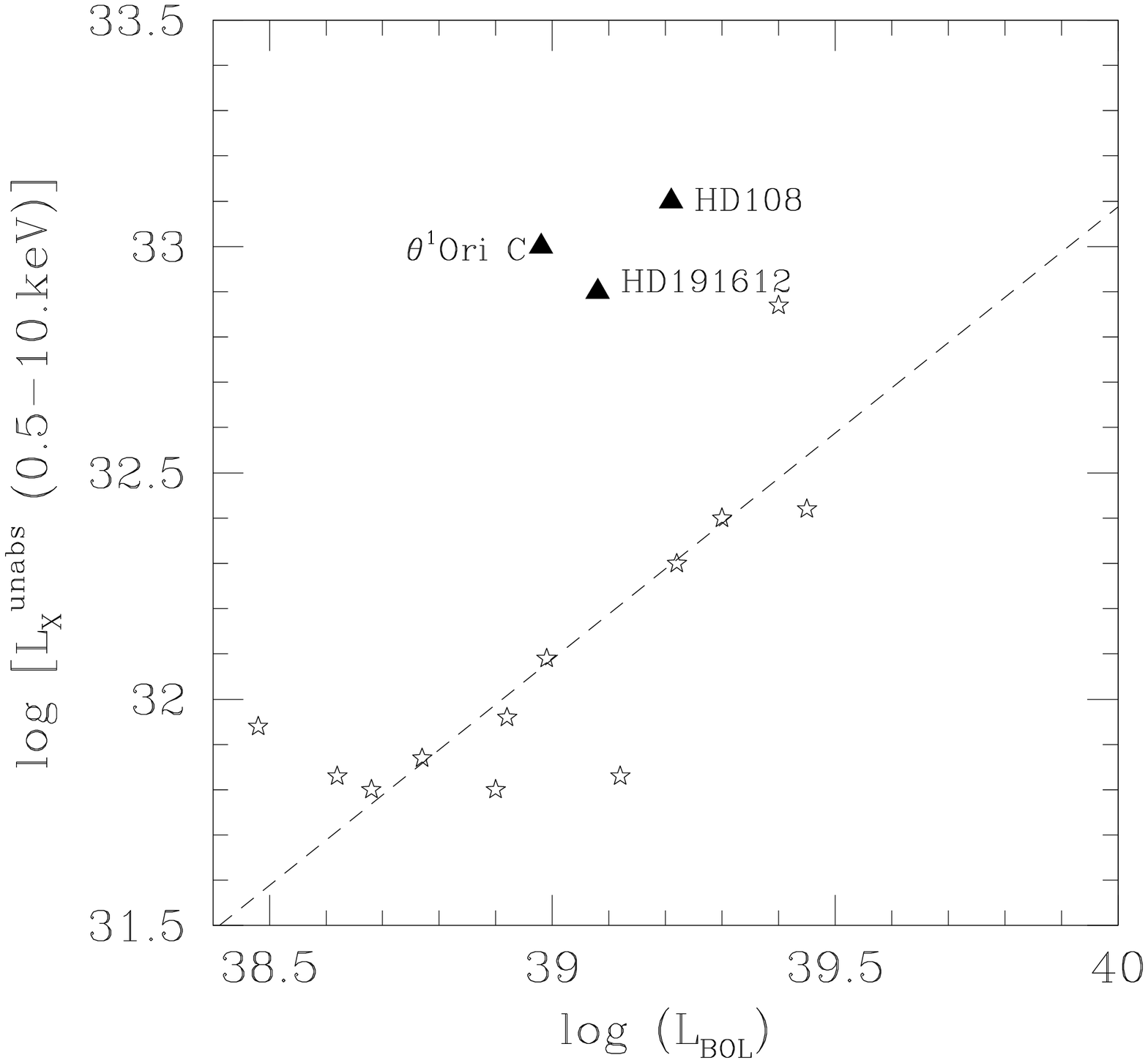}
\caption{Diagram showing the X-ray luminosity (in ergs s$^{-1}$) versus bolometric luminosity (in ergs s$^{-1}$). The dashed line indicates the typical relation for O stars (from \citealt{san06}), which HD\,108, \hd\ and \th\ all lie above. Asterisks show the position of hot stars in NGC\,6231 \citep{san06} with three outliers: the two objects lying above the line are colliding-wind binaries whereas the one lying below is a Wolf-Rayet binary. \label{lxlbol}}
\end{figure}

Using bolometric corrections from \citet{hum} and the optical properties of HD\,108 ($V$=7.4, $B-V$=0.18, $d$=2.51~kpc) and of \hd\ ($V$=7.8, $B-V$=0.27, $d$=2.29~kpc, \citealt{gies}), we calculated the bolometric luminosities of both objects: $L_{\rm BOL}$=1.6$\times 10^{39}$~erg~s$^{-1}$ for HD\,108 and 1.2$\times 10^{39}$~erg~s$^{-1}$ for \hd. The X-ray luminosities $L_{\rm X}^{\rm unabs}$, evaluated from the EPIC data in the 0.5--10.0\,keV range and corrected for interstellar absorption, are 1.2$\times 10^{33}$~erg~s$^{-1}$ for HD\,108 \citep{hd108xmm} and $7-9\times 10^{32}$~erg~s$^{-1}$ for \hd\ (this work). This results in $\log (L_{\rm X}^{\rm unabs}/L_{\rm BOL})$ ratios of $-$6.1 for HD\,108 and $-$6.1 to $-$6.2 for \hd. The values of the $L_{\rm X}^{\rm unabs}/L_{\rm BOL}$ ratio are thus five times larger than that of the `canonical' relation ($\log (L_{\rm X}^{\rm unabs}/L_{\rm BOL})=-$6.91 in the most recent determination of \citealt{san06}), as shown in Fig.~\ref{lxlbol}. This overluminosity is slight but significant, since the new study by \citet{san06} shows that the scatter around the `canonical' relation is only about 20\%.

The parameters found for \hd\ (two temperatures, soft/hard character, overluminosity) are very similar to those found for HD\,108 by \citet{hd108xmm}, confirming the similarities between these two Of?p stars, even in the X-ray domain.

\begin{table*}
\centering
\begin{minipage}{180mm}
\vspace{5mm}
\caption{Best-fitting models and X-ray fluxes at Earth for each \xmm\ observation of \hd. The fitted model has the form {\tt wabs($N_{\rm int}^{\rm H}$)*[wabs($N^{\rm H}_1$)*mekal(k$T_1$)+wabs($N^{\rm H}_2$)*mekal(k$T_2$)]}, with {\tt wabs($N_{\rm int}^{\rm H}$)}$=3.4\times10^{21}$~cm$^{-2}$. Quoted fluxes are in the 0.4$-$10.0\,keV energy range. The unabsorbed fluxes $f_{\rm X}^{\rm unabs}$ are corrected only for the interstellar absorbing column. For each parameter, the lower and upper limits of the 90\% confidence interval (derived from the {\sc error} command under XSPEC) are noted as indices and exponents, respectively. The normalisation factors are defined as $\frac{10^{-14}}{4\pi D^2} \int n_e n_{\rm H} dV$, where $D$, $n_e$ and $n_{\rm H}$ are respectively the distance to the source, the electron and proton density of the emitting plasma; $\phi$ is the phase in the 538\,d cycle.
\label{tab: table_fit}}
\setlength{\tabcolsep}{1.5mm}
\begin{center}
\begin{tabular}{ccccccccclcc}
\hline
Date & Rev. & $\phi$ & $N_1^{\rm H}$  & k$T_1$   & $norm_1$     & $N_2^{\rm H}$  & k$T_2$ & $norm_2$     & $\chi^2_{\rm \nu}$(dof) & $f_{\rm X}^{\rm abs}$ & $f_{\rm X}^{\rm unabs}$\\
&     & & $10^{22}$~cm$^{-2}$ & keV   & $10^{-3}$cm$^{-5}$         & $10^{22}$~cm$^{-2}$ & keV  & $10^{-3}$cm$^{-5}$        &                               & \multicolumn{2}{c}{($10^{-13}$~erg\,cm$^{-2}$\,s$^{-1}$)}  \\
\hline
\multicolumn{10}{l}{`cool' model}\\
\vspace*{-0.3cm}&&&&&&\\
05/04/05 & 975  & 0.09 & $0.51_{0.42}^{0.61}$ & $0.23_{0.20}^{0.25}$ & $7.07_{4.04}^{15.9}$ &  $1.07_{0.87}^{1.31}$ & $1.27_{1.19}^{1.36}$ & $1.08_{0.96}^{1.21}$ & 1.16 (411) & 7.0 & 14.1\\
\vspace*{-0.3cm}&&&&&&\\
17/04/05& 981  & 0.12 & $0.32_{0.25}^{0.46}$ & $0.27_{0.23}^{0.29}$ & $2.15_{1.40}^{3.81}$ & $0.91_{0.80}^{1.10}$ & $1.01_{0.96}^{1.07}$ & $1.22_{1.11}^{1.35}$ & 1.10 (681) & 6.8 & 14.3\\
\vspace*{-0.3cm}&&&&&&\\
02/06/05& 1004 & 0.20 & $0.43_{0.31}^{0.51}$ & $0.25_{0.23}^{0.28}$ & $3.78_{1.78}^{5.83}$ & $1.03_{0.81}^{1.25}$ & $1.22_{0.98}^{1.31}$ & $0.89_{0.74}^{1.11}$ & 1.15 (460) & 6.0 & 12.5\\
\vspace*{-0.3cm}&&&&&&\\
08/10/05& 1068 & 0.44 & $0.61_{0.58}^{0.64}$ & $0.20_{0.19}^{0.21}$ & $13.6_{10.7}^{18.1}$ & $1.39_{1.14}^{1.68}$ & $1.60_{1.49}^{1.72}$ & $0.66_{0.59}^{0.76}$ & 1.31 (555) & 5.3 & 11.0\\
\vspace*{-0.3cm}&&&&&&\\
\hline
\multicolumn{10}{l}{`hot' model}\\
\vspace*{-0.3cm}&&&&&&\\
05/04/05 & 975 & 0.09 & $0._{0.}^{0.05}$ & 0.64$_{0.62}^{0.66}$ & 0.28$_{0.27}^{0.33}$ & $0._{0.}^{0.03}$ & $2.35_{2.21}^{2.54}$ & $0.51_{0.48}^{0.53}$ & 1.19 (411) & 7.6 & 14.4\\
\vspace*{-0.3cm}&&&&&&\\
17/04/05& 981  & 0.12 & $0._{0.}^{0.01}$ & $0.60_{0.58}^{0.61}$ & $0.32_{0.31}^{0.34}$ & $0.81_{0.70}^{0.87}$ & $1.23_{1.06}^{1.32}$ & $0.95_{0.86}^{1.15}$ & 1.08 (681) & 7.0 & 13.9\\
\vspace*{-0.3cm}&&&&&&\\
02/06/05& 1004 & 0.20 & $0._{0.}^{0.01}$ & $0.51_{0.47}^{0.54}$ & $0.28_{0.26}^{0.30}$ & $0.82_{0.74}^{0.91}$ & $1.07_{1.01}^{1.27}$ & $0.92_{0.75}^{1.01}$ & 1.16 (460) & 5.8 & 12.0\\
\vspace*{-0.3cm}&&&&&&\\
08/10/05& 1068 & 0.44 & $0._{0.}^{0.01}$ & $0.61_{0.60}^{0.63}$ & $0.23_{0.22}^{0.24}$ & $0._{0.}^{0.01}$ & $2.53_{2.38}^{2.68}$ & $0.32_{0.31}^{0.34}$ & 1.39 (555) & 5.4 & 11.0\\
\vspace*{-0.3cm}&&&&&&\\
\hline
\end{tabular}
\end{center}
\end{minipage}
\end{table*}

\subsection{High-Resolution Spectroscopy}
Due to the weakness of \hd\ in X-rays, individual RGS spectra do not have enough counts for a sound analysis (see, however, below). We have therefore first combined the data from the three observations with the lowest background levels (i.e., Revs. 975, 1004, and 1068) using the task {\sc rgscombine} and then assembled the first and second-order spectra of both instruments into a single RGS spectrum by means of the {\sc rgsfluxer} command. 

\begin{figure}
\centering
\includegraphics[width=8cm]{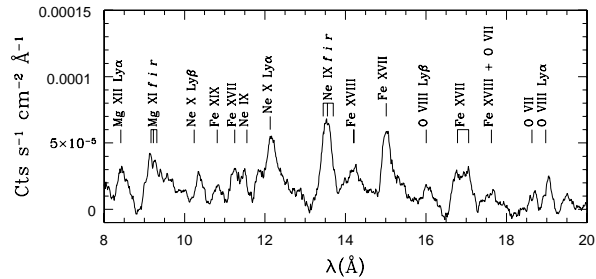}
\caption{Smoothed combined RGS spectrum of HD\,191612 between 8 and 20\,\AA\  (RGS1 + RGS2, first and second order, Revs. 975, 1004, and 1068). The strongest lines are identified.\label{rgsspec}}
\end{figure}

\begin{figure}
\centering
\includegraphics[width=8cm, bb=25 145 585 425, clip]{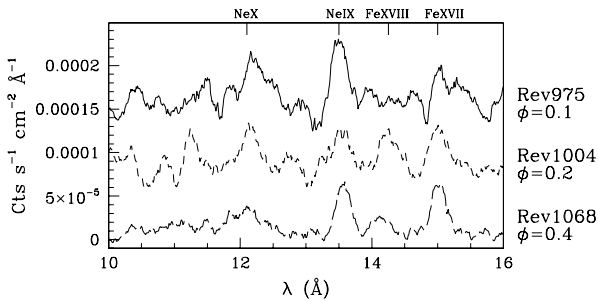}
\caption{Smoothed combined RGS spectra of HD\,191612 (RGS1 + RGS2, first and second order). The spectra from Rev.\ 1004 and 975 are shifted upwards by $7 \times 10^{-5}$ and $1.4 \times 10^{-4}$ units respectively. Note the changing intensity of the Fe\,{\sc xviii} lines at $\lambda\lambda$\,11.421, 14.204, 14.212 and Ne\,{\sc x} Ly$\alpha$  compared to the neighbouring lines of Ne\,{\sc ix} (helium-like triplet) and Fe\,{\sc xvii}.  \label{rgsvariations}}
\end{figure}

Figure\,\ref{rgsspec} shows the resulting RGS spectrum in the wavelength range 8--20\,\AA. Due to the rather heavy interstellar and circumstellar absorption, and the weakness of the source, the signal-to-noise ratio of RGS data above $\sim 20$\,\AA\ is too low to be useful. The spectrum of HD\,191612 is clearly dominated by emission lines and, apart from some higher excitation lines (e.g. Fe\,{\sc xviii}) that appear relatively strong here, it looks quite similar to the RGS spectrum of typical O stars such as $\zeta$\,Pup \citep{Kahn}. In particular, we note that the RGS spectra do not show a strong bremsstrahlung continuum emission unlike what is seen in the HETG spectra of the colliding-wind binary WR140 \citep{pol05} or the peculiar magnetic O star \th\ \citep{sch00}. 

We identified the strongest lines by comparison with the line list of the SPEX plasma code \citep{kaa}. The most prominent lines are hydrogen- and helium-like transitions from Mg, Ne, and O together with lines from Fe\,{\sc xvii} and Fe\,{\sc xviii} (see Fig.\,\ref{rgsspec}). We also note the possible presence of weak spectral signatures of additional iron ions; for example, the strongest Fe\,{\sc xix}  line ($\lambda$\,13.52) might be present, but would be heavily  blended with the strong Ne\,{\sc ix} He-like triplet. The  strongest Fe\,{\sc xx} lines could actually be responsible for the low-level features near 12.8\,\AA, whilst the strongest  Fe\,{\sc xxi} line ($\lambda$\,12.29) could account for some of  the breadth of the Ne\,{\sc x} Ly$\alpha$ line. The spectral signature of these additional ionization stages of iron is indeed expected since the EPIC spectra reveal the likely presence of an Fe XXV line at 6.7\,keV. The temperatures of maximum emissivity of the most prominent lines seen in the RGS spectra span a range from $2$ -- $3 \times 10^6$\,K (O\,{\sc vii} and O\,{\sc viii}) to $7$ -- $10 \times 10^{6}$\,K (Fe\,{\sc xviii} and Mg\,{\sc xii}). These values, in agreement with the results from the low-dispersion spectroscopy, are again very similar to those found for the Of?p star HD\,108 \citep{hd108xmm}.

The rather limited signal-to-noise ratio of the spectrum prevents us from doing a detailed profile analysis for individual lines. We have nevertheless measured the {\sc fwhm} of O\,{\sc viii} Ly$\alpha$ and the strong Fe\,{\sc xvii} $\lambda$\,15.014 line on the combined (unsmoothed) RGS spectrum, finding {\sc fwhm} values between 2000 and 2600\,\kms. We also tentatively performed a global fit of the X-ray lines by a single triangular\footnote{The triangular profile gives an unbiased estimate of any shift and asymmetry and is similar to several proposed theoretical profiles.} profile whose amplitude was allowed to vary from line to line but whose width and shift were taken to be the same for all lines. Within the limit of our RGS data, we detect no blue/red shift $>$250\,\kms nor any significant line asymmetry. The FWHM is found to be about 1800$\pm$400\,\kms, or 70-85\% of the terminal velocity \citep{wal03}, suggesting again rather broad lines. As the above values correspond to about twice the instrumental resolution, it seems that the spectral lines of HD\,191612 are indeed broad and can be resolved with the RGS. 

The strength of some of the spectral lines changes between the various \xmm\ pointings. For instance, the Fe\,{\sc xviii} lines appear strongest in the spectrum from June\ 2 (Rev.~1004, $\phi$=0.2), when the star is in an intermediate state: at that time, Fe\,{\sc xviii} $\lambda\lambda$\,14.212, 14.260 have an intensity comparable to that of Fe\,{\sc xvii} $\lambda$\,15.014. However, these lines are weaker in the first and last observations (Revs. 975 and 1068,  $\phi$=0.1 and 0.4; Fig.\,\ref{rgsvariations}). A similar difference is seen for the Ne\,{\sc x} and Ne\,{\sc ix} lines. These variations most probably reflect the changing hardness ratio measured from the EPIC data of HD\,191612. 

\begin{figure*}
\centering
\includegraphics[bb=18 490 590 700, clip, width=18cm]{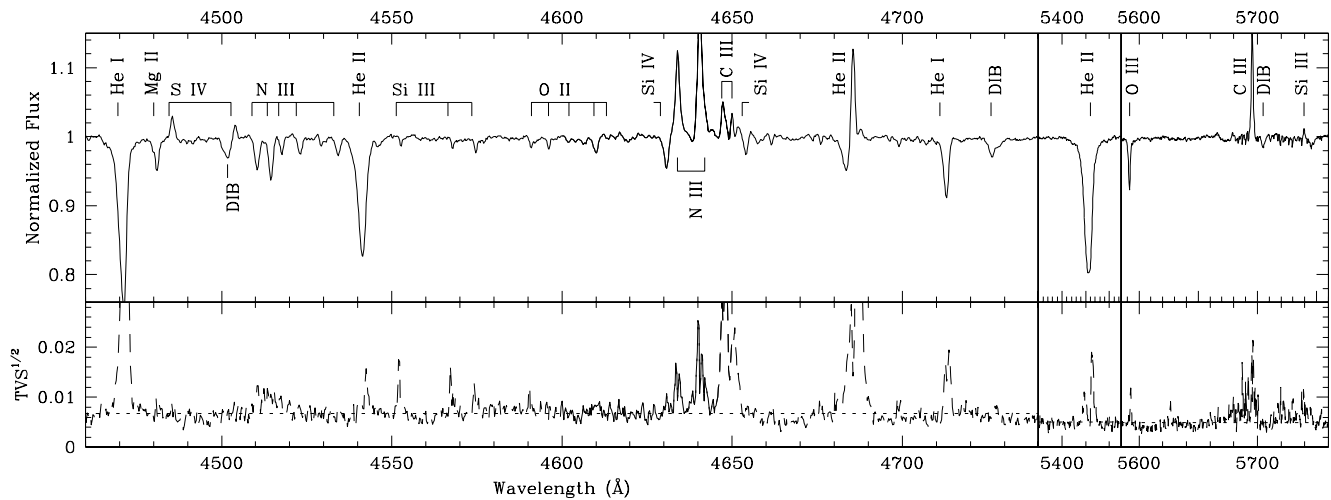}
\caption{\label{fig: spec} Mean spectrum of \hd\ (top panels) and the TVS (bottom panels). The main lines are indicated at their rest wavelengths. For the TVS panels, a dotted horizontal line indicates the significance level at 99\%. Some variability due to changing weather conditions appears at the position of the telluric lines, e.g. near \ciii\,\l\,5696. }
\end{figure*}

\begin{figure}
\centering
\includegraphics[bb=40 180 390 710, clip, width=4cm]{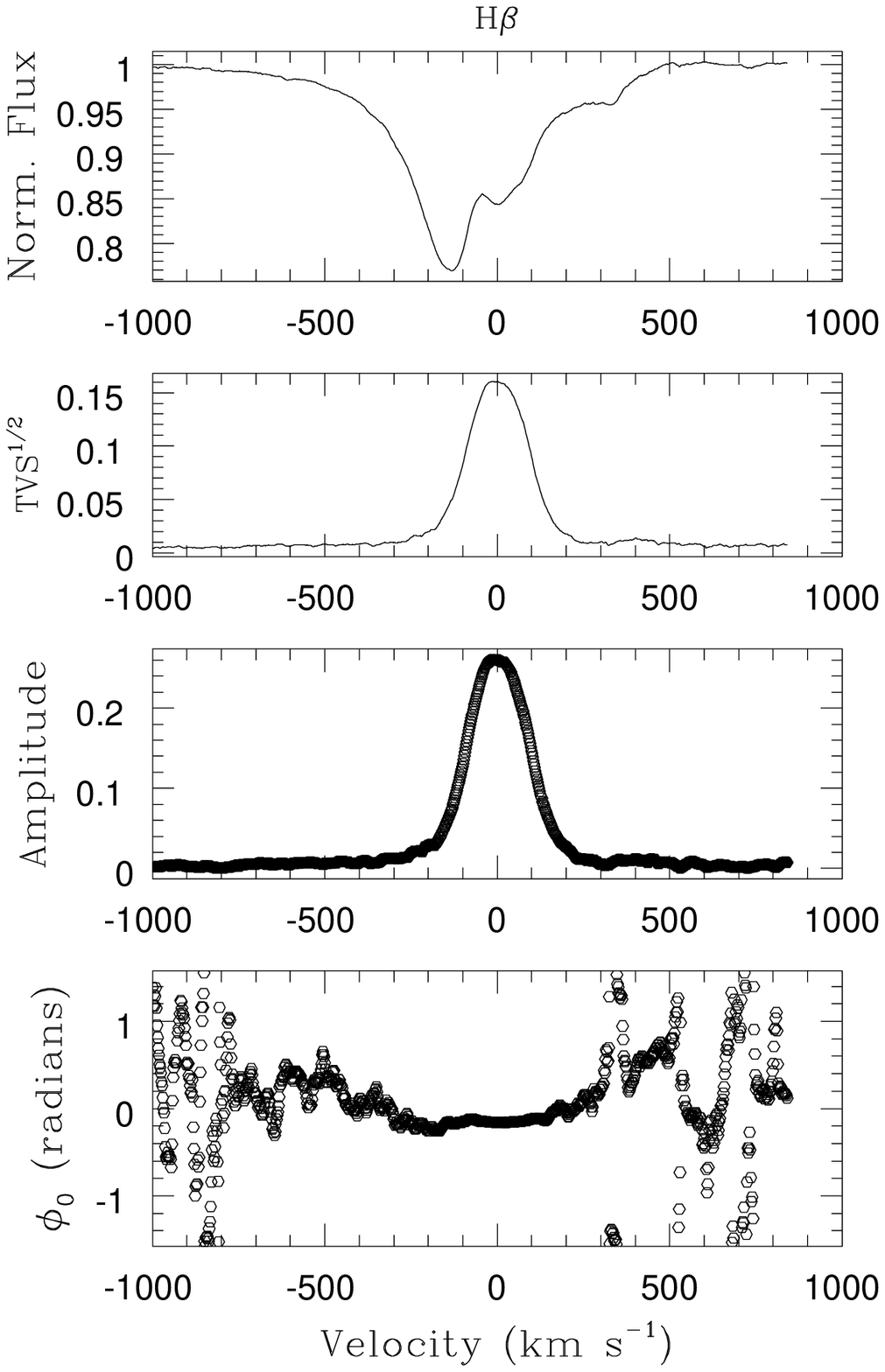}
\includegraphics[bb=40 180 390 710, clip, width=4cm]{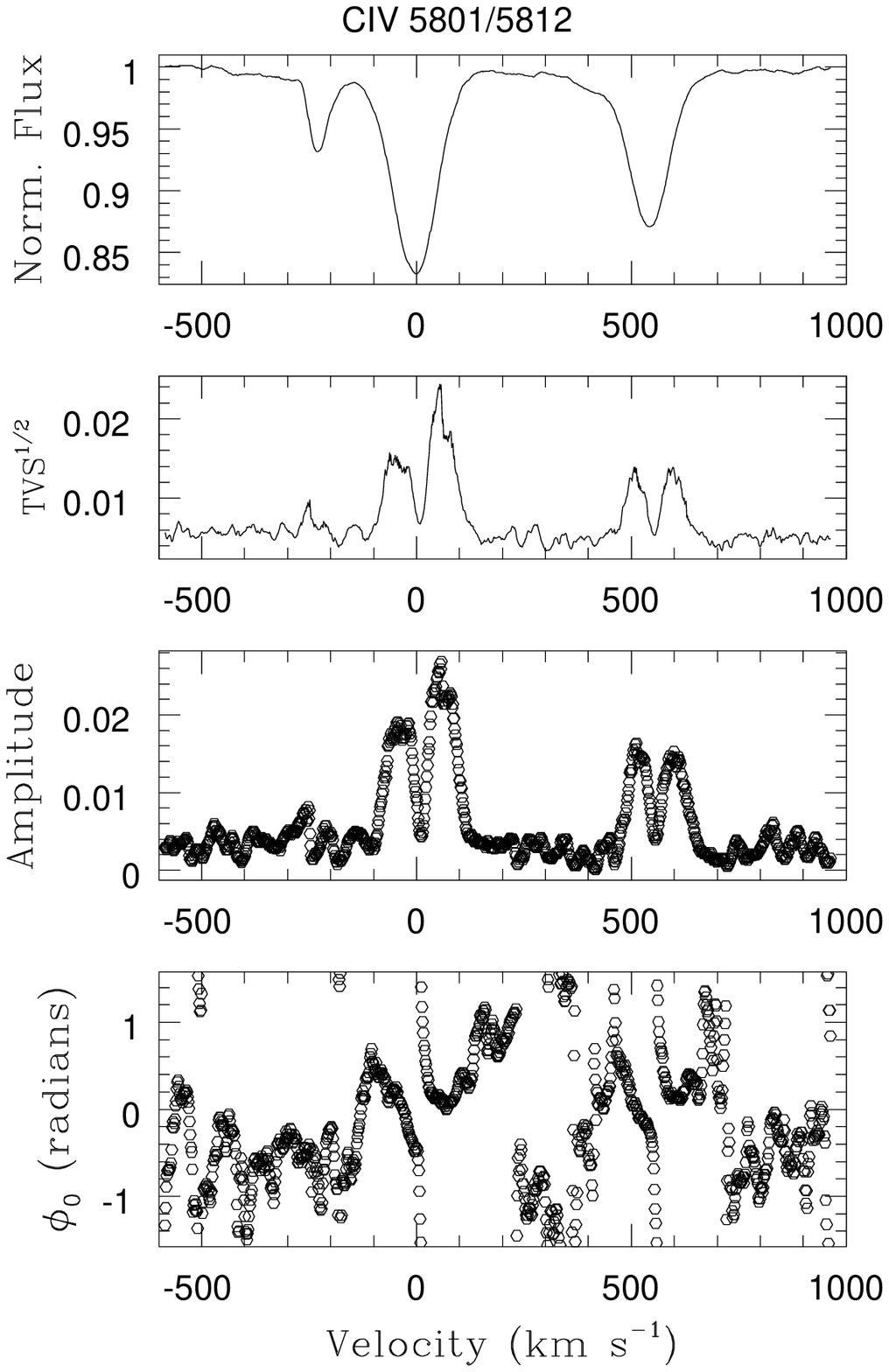}
\caption{\label{fig: ampl} Variations of \hb\ (left panel) and \civ\,\l\l\,5801,5812 lines (right panel). The individual panels display, from top to bottom, the mean line profile, the TVS, as well as the amplitude and the phase of the sine wave of frequency 1/537.7~d$^{-1}$ fitted to our data.}
\end{figure}

\section{Properties in the optical range}
To identify the lines displaying variability, we computed a Temporal Variance Spectrum (TVS, \citealt{ful}) for the Elodie data set (shown partly on Fig.~\ref{fig: spec}). As already mentioned in \citet{naz01} for HD\,108 and in \citet{wal03} for \hd, the most significant profile variations are found in the Balmer and \he\ lines. The TVS of these lines displays a roughly gaussian shape centered on the rest wavelength. It is important to note that the blue wings of the profiles, unaffected by the emission, exhibit very little variability (Fig.~\ref{fig: ampl}). The Balmer and \he\ lines thus consist of an apparently constant\footnote{Because of the presence of the varying emission, small RV variations as those exhibited by the metallic lines (see below) would not be detectable for the observed H and \he\ lines.} absorption component with a superimposed varying emission profile extending from approximately $-$200 to 200\,\kms. For O-type stars, a strong \ciii\,\l\l\,4649,4650 emission is quite rare. Its presence in \hd\ cannot be due to an abundance effect \citep{wal00} because the strength of these lines varies while \ciii\,\l\,5696 remains constant. Since the emission mechanism of \ciii\,\l\l\,4649,4650 is rather complex \citep{vreuxand}, the explanation of their variability requires detailed atmospheric modelling.

On the other hand, now that we have data with high signal-to-noise covering the whole 538\,d period, we see that almost all lines present significant changes (Fig.~\ref{fig: spec}). The variability characteristics, though, are different from those of the Balmer and \he\ lines since their TVS generally displays a double-peaked shape (Fig.~\ref{fig: ampl}).

We have analyzed further the spectral variations by using techniques similar to those of \citet{rauw}. At every wavelength step, we have fitted a sine wave of frequency 1/537.7~d$^{-1}$ to our time series. Fig.~\ref{fig: ampl} shows the amplitude and the phase of this sine wave as a function of radial velocity for two different spectral regions. Note that the phase $\phi_0$ has no physical meaning outside the wavelength domain where a significant variation is detected. This explains the noisy phases appearing at the extremes of our velocity interval. 

Once again, the two subsets of lines display different behaviours, in phase as well as in amplitude. Absorption lines from \hee\ and abs./em. lines from metals present a double-peaked amplitude profile and a varying phase, whereas for the Balmer and \he\ lines, the amplitude profile is gaussian and the phase is rather constant (Fig.~\ref{fig: ampl}). We note however that a small double-peaked feature is apparently superimposed on the gaussian amplitude profile for the weakest \he\ lines and \hee\,\l\,4686.

To further study the spectral behaviour of \hd, we decided to focus on the \civ\,\l\l\,5801,5812 lines since (1)~in our data, these lines belong to the domain presenting the highest signal-to-noise ratio and (2) they display a very clean gaussian shape that does not seem affected by any residual emission or distortion of the line profile. These lines are thus most likely to arise in the stellar photosphere, unlike the Balmer and \he\ lines which are often contaminated by variable wind emission. The results obtained from these \civ\ lines should therefore remain unaffected by strong noise or changing wind properties.

First, the equivalent widths (EWs) were measured by integrating the line profiles and we found that they remain rather stable, with an apparently stochastic variability of maximum 10\%. A similar result was found for other probable photospheric lines (i.e. metallic lines and \hee\ absorptions). In marked contrast, the variable hydrogen, \he\ and \hee\,\l\,4686 lines follow a very different behaviour, as was described in \citet{wal04} for \ha: their EWs first rise smoothly, reach a maximum emission level and then decline smoothly (Fig. \ref{fig: haxmm}).  

\begin{figure}
\centering
\includegraphics[width=4cm]{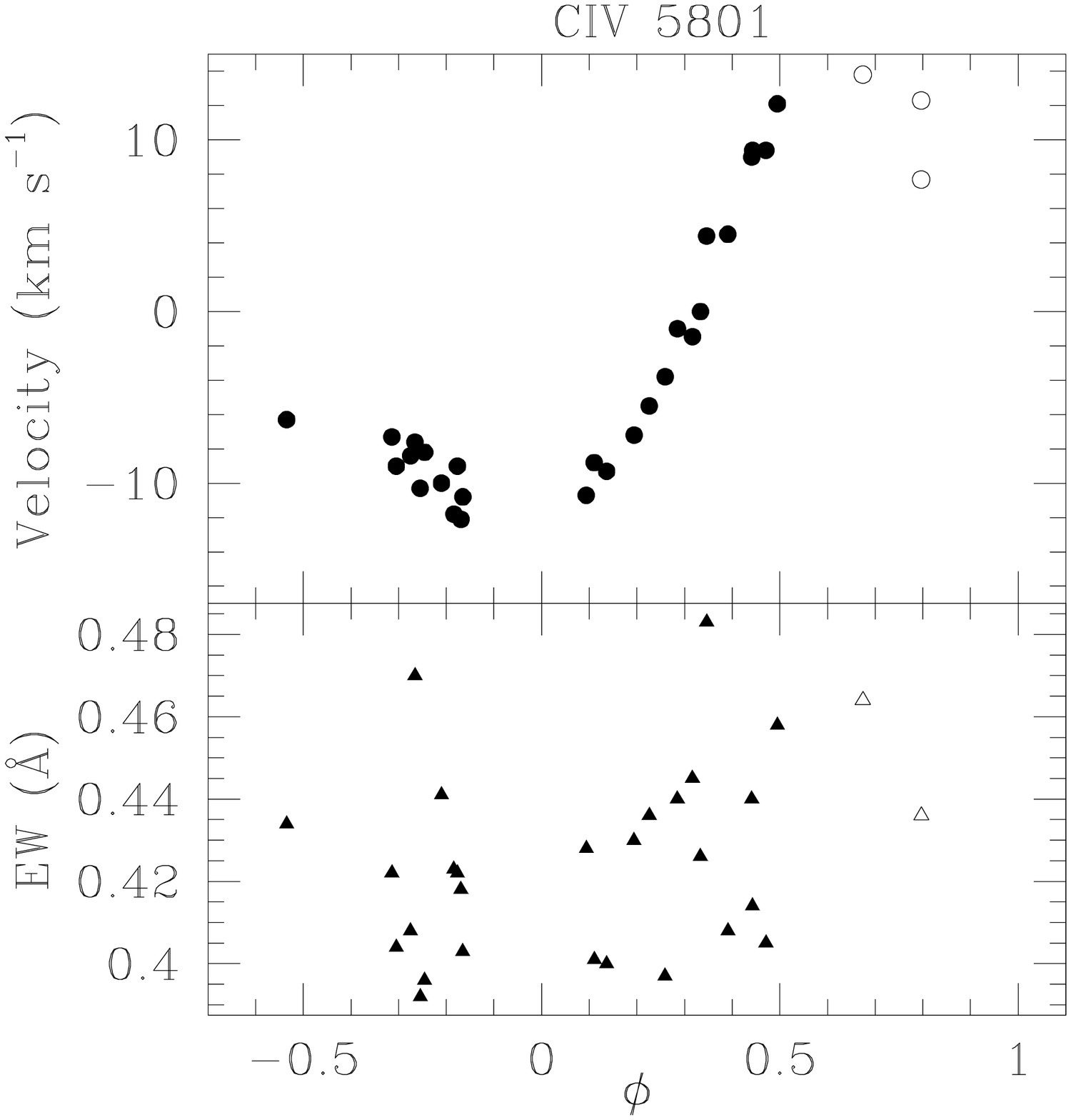}
\includegraphics[width=4cm]{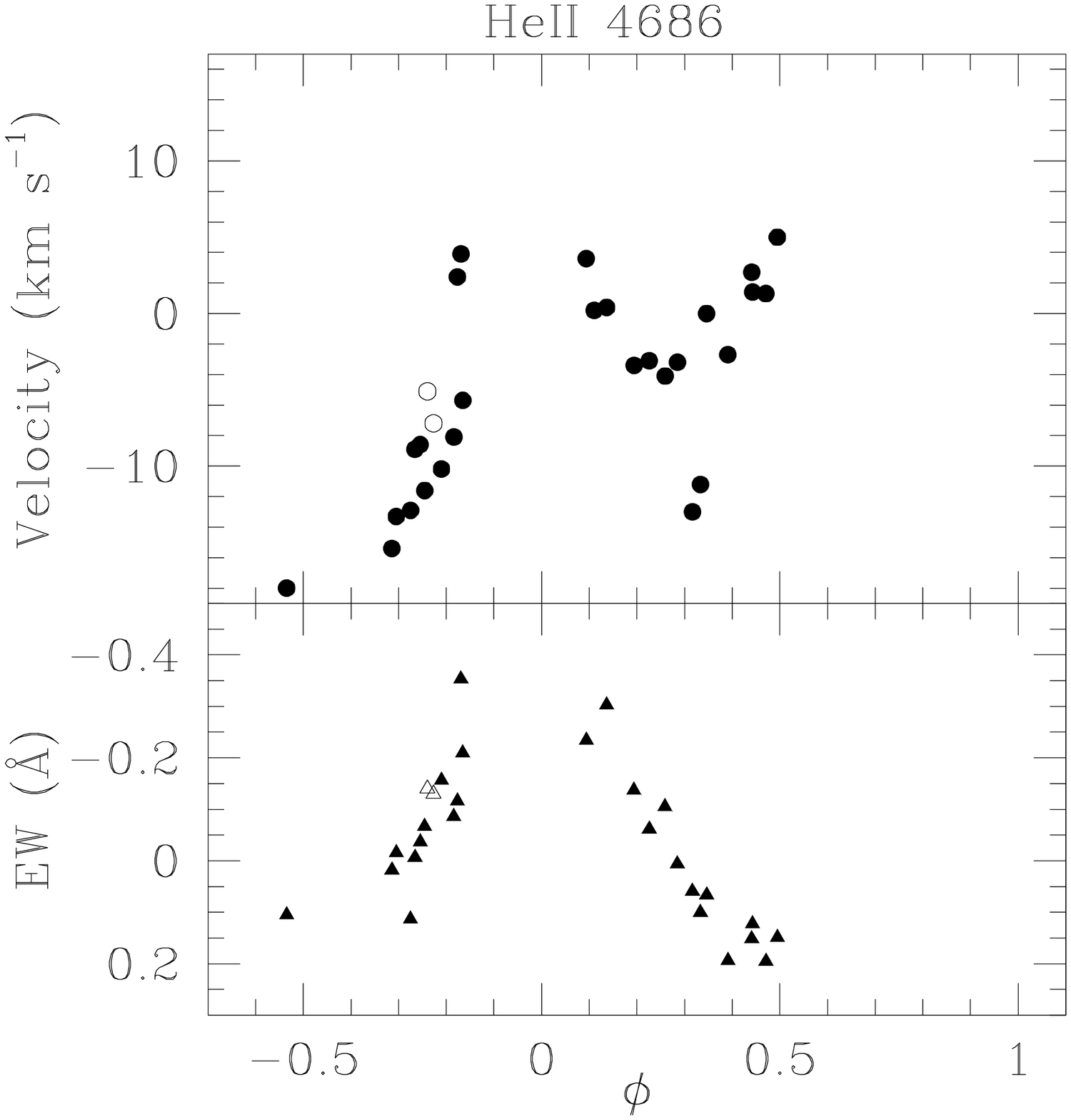}
\caption{\label{fig: rvew} {\it Left panel:} Radial velocity (top) and equivalent width (bottom) measurements for the \civ\,\l\,5801 line. Filled and open symbols stand for Elodie and Aur\'elie data, respectively. {\it Right panel:} Same for \hee\,\l\,4686. The RV points corresponds to the highest emission peak. Typical errors are 1--2\kms\ for RVs and 0.05\AA\ for EWs.}
\end{figure}

The radial velocities (RVs) were determined by fitting a gaussian to the \civ\ line profile. They appear systematically larger in the last OHP observations (Fig.~\ref{fig: rvew}). Again, similar results were found for the other strong metallic lines and the \hee\,\l\l\,4542,5412 absorption lines (where the RVs were derived from a gaussian fit to the bottom - resp. top - of the absorption - resp. emission - profile). 

Since the \civ\ lines are most probably of photospheric nature, their apparent motion can {\it a priori} be attributed to the star itself with some confidence. Although the periodic behaviour of these RVs cannot be assessed here with confidence, such a trend in RVs together with the double-peaked profile of their TVS suggest a binary nature for \hd\ (although probably with a period differing from 538\,d). 

The case of the \hee\,\l\,4686 emission line is peculiar: the RV of its emission peak follows the same trend as the RVs of metals when the emission is minimum (Fig.~\ref{fig: rvew}). This suggests that there are two components to the emission of \hee\,\l\,4686, one varies in harmony with the Balmer lines (and thus represents an additional emission) while the other simply shifts in wavelength as for the metallic lines (thus probably belonging to the star). Note that an additional component, in absorption, is also clearly detected. This might indicate the presence of a companion.

\section{Discussion}
After presenting the detailed analysis of the phase-resolved observations taken in the X-ray and optical ranges, we now discuss the implications of these results on the possible nature of \hd.

\subsection{\hd, a magnetic oblique rotator?}
Because of the detection of a magnetic field, one may think of \hd\ as an analogue to the other two magnetic hot stars \th\ and $\tau$ Sco (see e.g. \citealt{donati}). Indeed, these systems share several similarities (see Table \ref{tab: compa_magn} for a comparison). For example, they display large magnetic confinement parameters ($\eta=\frac{B^2_{eq}R^2_*}{\dot M v_{\infty}}$) and their Alfv\'en radii are also similar (about 2~R$_*$; \citealt{donati,don06}). In addition, their $\log (L_{\rm X}^{\rm unabs}/L_{\rm BOL})$ ratios, $-$6.0 for \th\ and up to $-$6.1 for \hd, indicate an important overluminosity in X-rays compared to `typical' O stars (Fig. \ref{lxlbol}). Finally, these three magnetic objects display variations in the X-rays that seem correlated to those observed in the visible domain \citep[this work]{Gagne}. 

However, our new data also reveal many differences. First of all, the X-ray spectra of \th\ and $\tau$ Sco are much harder than that of \hd. The temperatures of the hot plasma of \th\ amount to 0.7 and 2.6\,keV \citep[i.e. 8~and~30MK; ][]{Gagne} where the higher-temperature component clearly dominates. The opposite situation is found for \hd, {\em even} in the case of the `hot' fit. This distinct behaviour cannot be attributed to extrinsic causes since both systems share similar interstellar absorption: the hard X-ray emission of \hd\ is thus unlikely to be hidden by a high extrinsic column density. 

Differences are also evident when considering $\tau$ Sco, which has a somewhat softer spectrum than \th. In this case, the differences can be best appreciated by comparing the differential emission-measure fit given by \citet{mew03} with ours: again, the hard component is much reduced for \hd. The `hard' character of the X-ray emission of \th\ and $\tau$ Sco is also revealed by the presence of several strong, high-ionization lines (e.g. Fe\,{\sc xxiv}) which are weak or absent in the spectrum of \hd. 

We also note that \th\ presents a prominent bremsstrahlung continuum in its X-ray spectrum which is a clear indication that the emission-measure distribution is dominated by high temperatures. No such feature is seen in the case of \hd.

In addition, the X-ray emission of \th\ and $\tau$ Sco consists of very narrow lines (FWHM well below 1000\,\kms; see \citealt{Gagne,coh03}), as predicted theoretically for a magnetically confined wind, but this is not the case for \hd, as was shown in Sect. 3.3. Even the broadest lines of \th\ and $\tau$ Sco (notably O\,{\sc viii} Ly$\alpha$ and the strong Fe\,{\sc xvii} $\lambda$\,15.014 line) are still narrower than the lines of \hd. The line width measured for \hd\ is actually much more similar to that determined for `normal' O stars. In many respects, \hd\ thus appears as an intermediate case between the magnetic objects and the typical O stars  (Fig. \ref{compax}).

Finally, we note that the behaviour of \hd\ in the visible range also differs somewhat from that of \th, although it is difficult to compare them in detail. In fact, while the spectral variability of \hd\ is now well known, this is not always the case for \th: for example, variations of the \he\,\l\,4471/\hee\,\l\,4542 ratio of \th\ were reported by \citet{wal81} but the same ratio was subsequently found constant \citep{stahl}.

\begin{table*}
\centering
\begin{minipage}{160mm}
\vspace{5mm}
\caption{Properties of \th\ (from \citealt{Gagne}), \hd\ (from \citealt{wal03, donati}), and $\tau$\,Sco \ (from \citealt{don06}). The wind density $\rho_w$ at $v_{\infty}/2$ was calculated assuming a $\beta$-type law (with $\beta$=0.8) and is shown in units of 10$^{-14}$ g cm$^{-3}$. The X-ray luminosity corresponds to the 0.5--10.0~keV band. \label{tab: compa_magn}}
\setlength{\tabcolsep}{1.5mm}
\begin{center}
\begin{tabular}{lcccccccccc}
\hline
Object & sp. type & $R_*$ & $T_{eff}$ & \.M & $v_{\infty}$ & $\rho_w$  & $B_{dip}$ & $\eta$ & $L_X^{unabs}$& Period \\
 &  & ($R_{\odot}$) & (kK) & ($M_{\odot}$ yr $^{-1}$) & (\kms) &   & (G) & & (erg s$^{-1}$) & (d)\\
\hline
\hd & O6 & 16.7 & 37 & 6.1$\times 10^{-6}$ & 2600 & 5.9 & $-$1500 & 7.6 & 7--9$\times 10^{32}$ & 538\\
    & O8 & 18.7 & 35 & 2.5$\times 10^{-6}$ & 2100 & 2.4 &  &  29.&  & \\
\th & O5.5V hot & 8.3 & 45 & 1.4$\times 10^{-6}$ & 2980 & 4.7 & 1060 & 3.6 & 0.8--1.1$\times 10^{33}$ & 15\\
    & O5.5V cool & 9.1 & 42 & 5.5$\times 10^{-7}$ & 2760 & 1.7 &  & 12 &  & \\
$\tau$\,Sco & B0V & 5.2 & 32 & 2$\times 10^{-8}$ & 2000 & 0.3 & not dip. & $\sim$40 & 2--3$\times 10^{31}$ & 41\\
\hline
\end{tabular}
\end{center}
\end{minipage}
\end{table*}

The differences between \hd\ and \th\ or $\tau$ Sco, especially in the X-ray domain, are quite puzzling, in view of other similarities. One could imagine that the larger mass-loss rate of \hd\ compared to \th\ would prevent us from seeing the X-ray emission from material in the magnetosphere (i.e. below the Alfv\'en radius). However, recent analyses of the X-ray line profiles of $\zeta$ Pup \citep{kra03} and $\zeta$ Ori \citep{coh06} indicate that in these stars (which are more similar to \hd\ as far as $\dot M$ is concerned), X-ray emission from down to 1.5~$R_*$ can be observed. Therefore, if there is a magnetically confined wind region in \hd, we would expect to be able to see its X-ray emission, unless the actual magnetic field of \hd\ is much lower than the dipolar one (1.5kG) inferred by \citet{donati}. This is indeed possible since the exact magnetic configuration of \hd\ is not yet known (only one phase was observed by Donati et al.). The discrepancies found between the actual properties of \hd\ and the magnetic oblique rotator model (as represented by \th) indicate that the emission mechanisms of these peculiar stars is not fully understood yet.

\subsection{Alternative Scenarios}
Since there are still problems with the magnetic oblique rotator scenario, we might envisage alternative models. If we assume that the RV variations are periodic and indeed provide the signature of a binary, the low amplitude of the RV curve (30\,\kms) would suggest a low mass function, which leaves us with two possibilities. 

First, that the system of \hd\ is seen at a low inclination and is then composed of two `normal' early-type stars. In this case, one may immediately think of colliding-wind (CW) phenomena. Indeed, \citet{san06} found that large $\log (L_{\rm X}^{\rm unabs}/L_{\rm BOL})$ ratios were displayed by the only two CW systems in their sample (see Fig. \ref{lxlbol}). In addition, the presence of a high-temperature component in the X-ray spectrum, and the modulations of the optical and X-ray emissions, are generally considered as typical signatures of this phenomenon. However, a detailed hydrodynamical simulation would be required to predict the putative colliding-wind emissions and their variations; while further X-ray observations are necessary to define better the correlations between X-ray and optical variations. In addition, no obvious signature of a massive companion has yet been detected in optical observations, except maybe for the peculiar shape of the \hee\,\l\,4686 profile (but see Howarth et al., in prep.).

On the other hand, \hd\ could be a high-inclination system with a lower-mass companion. Owing to the youth of \hd, this companion, if stellar, should still be pre-main-sequence (PMS) but the low luminosity of such objects \citep[$M_V\sim-2.5$ for a 7M$_{\odot}$ star, ][]{siess} compared to mid-O-type stars could prevent its emission being seen in the composite spectrum of the system. The alternative case of an accreting, compact object seems more promising. Of course, high accretion rates can be readily ruled out since they would lead to a much larger X-ray luminosity (see \citealt{hd108xmm} and references therein). However, the observed slight overluminosity in X-rays might be compatible with low to moderate accretion rates. Variations of the optical and X-ray emissions would be explained by the eccentricity of the orbit. However, this model also encounters difficulties when analyzed in detail: for example, such an eccentric orbit would result in asymmetric EW curves for \ha\ and similar lines, which is not the case. 

\section{Conclusions}
We have obtained phase-resolved observations of \hd\ with \xmm\ and the Elodie/Aur\'elie spectrographs of the OHP. 

\begin{figure}
\centering
\includegraphics[height=8cm, angle=90]{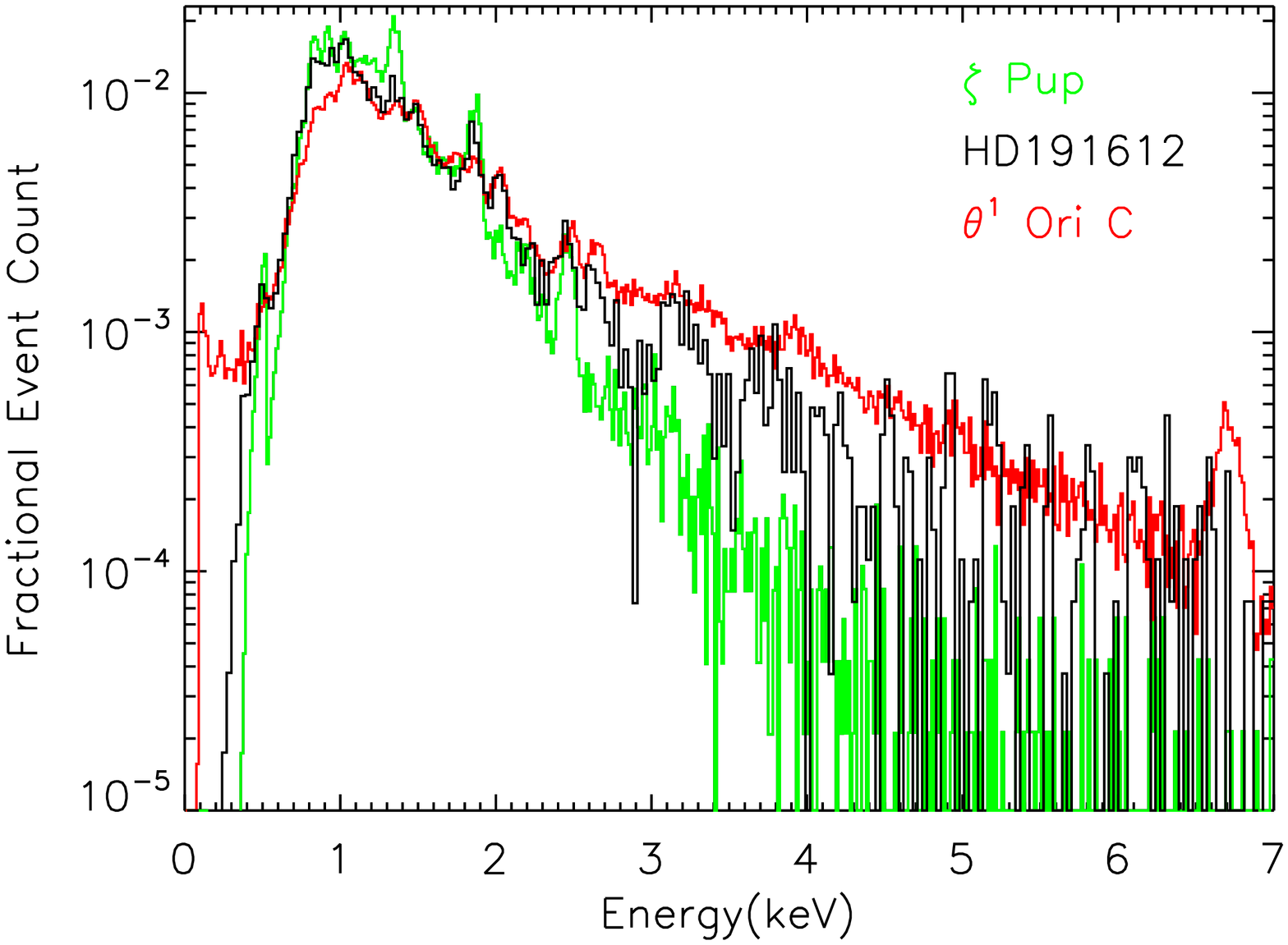}
\caption{EPIC MOS spectra of $\zeta$ Pup, \hd, and \th: \hd\ appears as an intermediate case between the ``typical'' O-type star (soft spectrum) and the oblique magnetic rotator (hard spectrum). The spectra have been arbitrarily re-normalized to appear with similar strengths, and an extra absorption was added to have all stars under $N_{\rm H}$=4.5$\times10^{21}$\,cm$^{-2}$, therefore correcting for the dissimilar interstellar absorptions of these objects ($log(N_{\rm H})$ of 20.0, 21.5, and 21.7 for $\zeta$ Pup, \hd, and \th, respectively). \label{compax}}
\end{figure}

In the high-energy domain, the star presents the spectrum of thermal plasmas with temperatures of 0.2--0.6 and 1--2\,keV. The cooler component dominates, and no strong lines with ionization higher than Fe\,{\sc xviii} are observed. The width of the X-ray lines, measured in the high-resolution RGS spectra, is 1800$\pm$400\,\kms. These results, which are similar to those found for `normal' O stars, are at odds with the predictions of a simple magnetic rotator model. Although no variability is detected during each pointing, the X-ray flux decreased by 40\%  between the first and last \xmm\ observations which were taken during high and low emission states (as defined in the visible), respectively. The high-energy variations might thus follow the optical changes. The average X-ray flux exceeds by a factor of 5 that derived from the `canonical' $L_{\rm X}^{\rm unabs}/L_{\rm BOL}$ relation. 

A clear dichotomy is seen in the behaviour of the optical lines. On the one hand, the hydrogen and \he\ lines display a relatively constant absorption superimposed on an emission of varying intensity (centred on the rest wavelength). On the other hand, our high quality dataset enabled us to detect for the first time the variations of the \hee\ absorptions and metallic (abs. or em.) lines. This variability is apparently associated with shifts in radial velocity (up to 30\,\kms) which are reminiscent of a binary motion.

The \hee\,\l\,4686 line displays a mixed behaviour which could tentatively be explained by the presence of three components: an emission following the behaviour of the H and \he\ lines, an emission shifting in RV as the metallic lines, and one additional absorption, probably of stellar origin. 

With these new data, the mystery deepens on the nature of Of?p stars: HD\,108 and \hd\ display some properties of typical O-type stars but also share some similarities with peculiar objects such as \th\ (Fig. \ref{compax}). The puzzling results from our multiwavelength campaign thus challenge the models proposed to explain the behaviour of \hd. 

\section*{Acknowledgments}

We are grateful to the OHP observers who have taken data for us in the framework of our Elodie service monitoring campaign as well as to Micha\"el De Becker for taking some of the Aur\'elie data. The Li\`ege team acknowledges support from the Fonds National de la Recherche Scientifique (Belgium) and the PRODEX XMM and Integral contracts. Part of the OHP observations have been funded by the OPTICON access programme supported by the European Commission's 6th Framework Programme. This research is also partly supported by contracts P5/36 ``P\^ole d'Attraction Interuniversitaire'' (Belgian Federal Science Policy Office). \\

\label{lastpage}

\end{document}